\documentclass[modern]{aastex62}
\usepackage{amsmath}

\received{10-16-2019}
\accepted{10-23-2019}
\submitjournal{ApJ Letters}

\shorttitle{Secondary Infall in the  Seyfert's Sextet}
\shortauthors{L\'opez-Cruz et al.}
\watermark{Accepted for publication in the Astrophysical Journal Letters}
\setwatermarkfontsize{20 pt}

\begin{document}

\title{Secondary Infall in the Seyfert's Sextet: A Plausible Way Out
  of the Short Crossing Time Paradox} 

\author[0000-0002-1381-7437]{Omar L\'opez-Cruz}\email{omarlx@inaoep.mx}
\altaffiliation{2015-2016 Oliver L.  Benediktson Endowed Chair in
  Astrophysics, Department of Physics and Astrophysics, University of
  North Dakota, Grand Forks, ND 58202, USA} 
\affiliation{Coordiaci\'on de Astrof{\'\i}sica, Instituto Nacional de
  Astrof{\'\i}sica \'Optica y Electr\'onica (INAOE), Luis E. Erro
  No. 1, Sta. Ma. Tonantzintla, Puebla, C.P. 72840 M\'exico} 
\correspondingauthor{Omar L\'opez-Cruz}

\author[0000-0002-9790-6313]{H\'ector Javier Ibarra-Medel}
\affiliation{University of Illinois Urbana-Champaign, Department of
  Astronomy. 103 Astronomy Bulding, 1002 W Green St, Urbana IL 61801,
  U.S.A.} 
\affiliation{Instituto de Astronom{\'\i}a, Universidad Nacional
  Aut\'onoma de M\'exico (IA-UNAM), Cd. de M\'exico, M\'exico} 


\author[0000-0002-9790-6313]{Sebasti\'an F.  S\'anchez}
\affiliation{Instituto de Astronom{\'\i}a, Universidad Nacional
  Aut\'onoma de M\'exico (IA-UNAM), Cd. de M\'exico, M\'exico} 

\author[0000-0002-1858-277X]{Mark Birkinshaw}
\affiliation{HH Wills Physics Laboratory, University of Bristol,
  Tyndall Avenue, Bristol, BS8 1TL, UK} 


\author[0000-0002-3721-8869]{Christopher A\~norve}
\affiliation{Facultad de Ciencias de la Tierra y el Espacio (FACITE),
  Universidad  Aut\'onoma de Sinaloa (UAS), Culiac\'an, Sinaloa,
  M\'exico} 

\author[0000-0003-2405-7258]{Jorge K. Barrera-Ballesteros}
\affiliation{Instituto de Astronom{\'\i}a, Universidad Nacional
  Aut\'onoma de M\'exico (IA-UNAM), Cd. de M\'exico, M\'exico} 

\author[0000-0002-0608-9574]{Jes\'us  Falcon-Barroso}
\affiliation{Instituto de Astrof{\'\i}sica de Canarias, Calle V{\'\i}a
  L\'actea s/n, E-38205, Espa\~na}
\affiliation{Departamento de Astrof{\'\i}sica, Universidad de La
  Laguna (ULL), E-38200 La Laguna, Tenerife, Espa\~na} 

\author[0000-0001-5547-3938]{Wayne A. Barkhouse}
\affiliation{Department of Physics and Astrophysics, University of
  North Dakota, Grand Forks, ND 58202, USA} 

\author[0000-0002-8009-0637]{Juan P. Torres-Papaqui}
\affiliation{Departamento de Astronom{\'\i}a, Universidad de
  Guanajuato (DAUG), Callej\'o—n Jalisco s/n Col. Valenciana,
  C.P. 36240 Guanajuato, Gto., M\'exico} 




\begin{abstract}
 We used integral field spectroscopy from CALIFA DR3 and
 multiwavelength publicly-available data to investigate the
 star-formation histories of galaxies in the Seyfert's Sextet (SS,
 HCG~79). The galaxies H79a, H79b, H79c, and H79f have low star-formation
 rates despite showing strong signs of interaction. By exploring their
 individual specific star formation histories (sSFH), we identified
 three earlier episodes of strong star formation common to these four
 galaxies. We use the last two episodes as markers of the epochs when
 the galaxies were crossing. We suggest that after the first
 turn-around, initially gas-rich galaxies crossed for the first time,
 consuming most of their gas. Hence after the second turn-around most
 mergers from second crossings would be mixed or dry. The exception
 would be gas-rich galaxies intruding for the first time. Therefore,
 we suggest that SS galaxies have survived one crossing during a
 Hubble time. Strong Balmer absorption lines and the presence of 
 counter-rotating disks provide independent bounds to the second and
 first crossing, respectively. This scenario provides a plausible way
 out of the short crossing time paradox.
\end{abstract}




\section{Introduction} \label{sec:intro}
For quite some time, it has been known that the crossing times
($t_{c}\thicksim{D}/{\sigma}$, where $D$ is the group's diameter and
$\sigma$ is the velocity dispersion) in compact groups (CG) are much
shorter than the Hubble time, $t_{H} = 12.6 \, h_{75}^{-1}$ Gyr 
\citep[e.g.,][]{H93}.  For example, \citet{Di93} used observations and
numerical simulations to find that $t_{c} \thicksim0.13\,t_{H}$.
 This, along with similar previous results, has led to the so-called
short dynamical time paradox \citep[e.g.,][]{W89,Di93,Ba97}. Naively,
one would, expect that most CG should have already collapsed into a
single merger remanent, resembling, perhaps, a giant early-type
galaxy. Many scenarios have been proposed as ways out of this apparent
paradox. Here, we consider the three most commonly called  scenarios: 
\begin{enumerate}
 \item CG are chance projections along the line-of-sight;
 \item CG have just assembled; and
 \item galaxies in CG have survived many crossings. 
\end{enumerate}
Scenario~2 is complementary to the replenishing scenario proposed by
\citet{Di93}. Below, we argue in favor of the third scenario. We
propose that CG have collapsed at least once during a Hubble time, but 
nonetheless, some group galaxies survived intact.

 
We have revisited \object{Seyfert's Sextet} 
\citep[SS, \object{HCG 79},][]{Se51}, a much-studied CG 
\citep[e.g.,][]{Pe71,Dur08}.  
The galaxies \object{H79a}, \object{H79b}, 
\object{H79c}, \object{H79d} and \object{H79f} (also a galaxy, 
see \S\ref{sSFH}) are members of SS, while \object{H79e} 
($z=0.06$) is a background galaxy. The naming of the galaxies follows 
\citet[][hereafter Du08]{H93, Dur08}. SS was singled out by
\citet{Pe71}, remarking its high galaxy density and short dynamical time. \citeauthor{Pe71}
went further by suggesting that spiral galaxies would be short-lived
in CG environments.  We can use SDSS data to show that SS is an an
isolated group \citep[e.g.,][]{DZ15}; hence, it's an excellent site to investigate how its
spiral galaxies have survived.

In this letter, we present an analysis of integral field spectroscopy
(IFS) observations from the Calar Alto Legacy Integral Field Area
Survey Data Release 3 (CALIFA DR3) \citep{Sa16a}, supplemented with
archival data from HST, SDSS, Spitzer, Herschel, and WISE
(\S\ref{obs}).  We have considered the fossil stellar record
(\S\ref{starlight}) of galaxies and secondary infall (see 
 \S\ref{model}) to develop a simple timing argument 
(\S\ref{sSFH}). The state of the gas, dust, and galactic activity and
dynamics of SS galaxies is presented in (\S\ref{dis}) to establish
the evolutionary stage of Seyfert's Sextet. Finally in \S\ref{con}, we
compare the three scenarios presented above, and summarize the results
and inferences drawn from this study. 

We assumed $H_0=75\,h_{75} \: \mathrm{km\, s^{-1}Mpc^{-1} }$, 
$\Omega_m=0.3$, and $\Omega_{\Lambda}=0.7$ throughout this paper,
to allow direct comparisons with previous works.

\begin{figure*}[t]
 \includegraphics[scale=.50,angle=0]{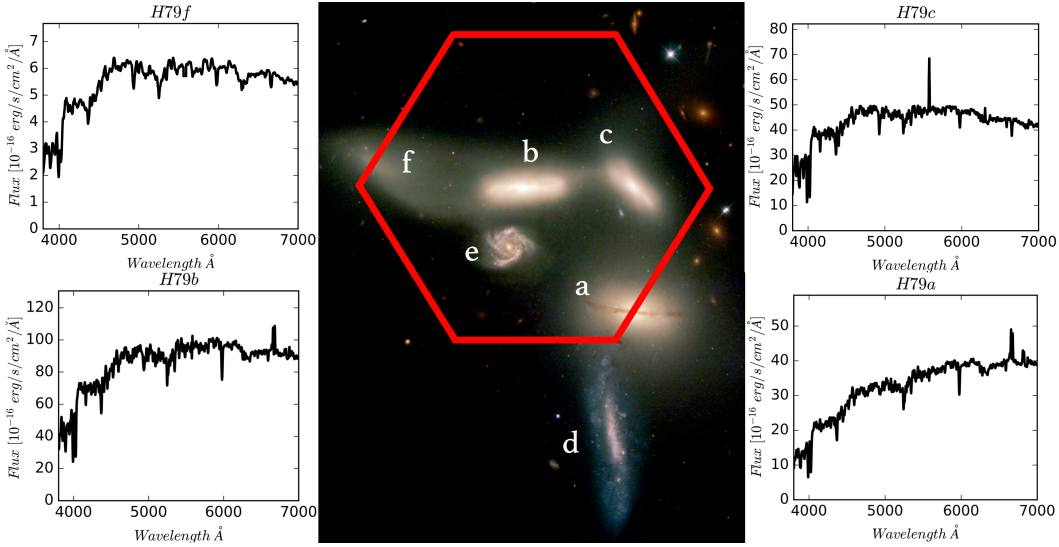}
 \caption{ 
  CALIFA's footprint, indicated by the red hexagon, on an HST composite
  image of Seyfert's Sextet, with the spectra of galaxies H79a, H79b,
  H79c, and H79f integrated down to $1.5 \times$ the effective radius,
  $R_{e}$. Spiral galaxy H79d is also a member of this CG, but was missed 
  by this HST pointing. H79e is a background galaxy. The wavelength
  axis (\AA) is indicated, while the spectral flux is in units of 
  $10^{-16}\; \mathrm{erg\,s^{-1}\,cm^{-2}\,\mathrm{\AA^{-1}}}$. 
  The $4000\, \mathrm {\AA}$ break and the absorption lines Mgb 
  $\lambda 5175$ and NaD $\lambda5892$ are clearly seen. Emission in
  $\mathrm{H}{\alpha}$ is seen in H79a and H79b. $\mathrm{H}{\beta}$ is
  seen in H79c. The color was composed from WFPC2 images using the
  filters F336W, F439W, F555W, and F814W --- see
  \url{https://hubblesite.org/image/1242/gallery} for further details.
 } 
 \label{fig1}
\end{figure*}

\begin{figure*}[t]
 \gridline{\fig{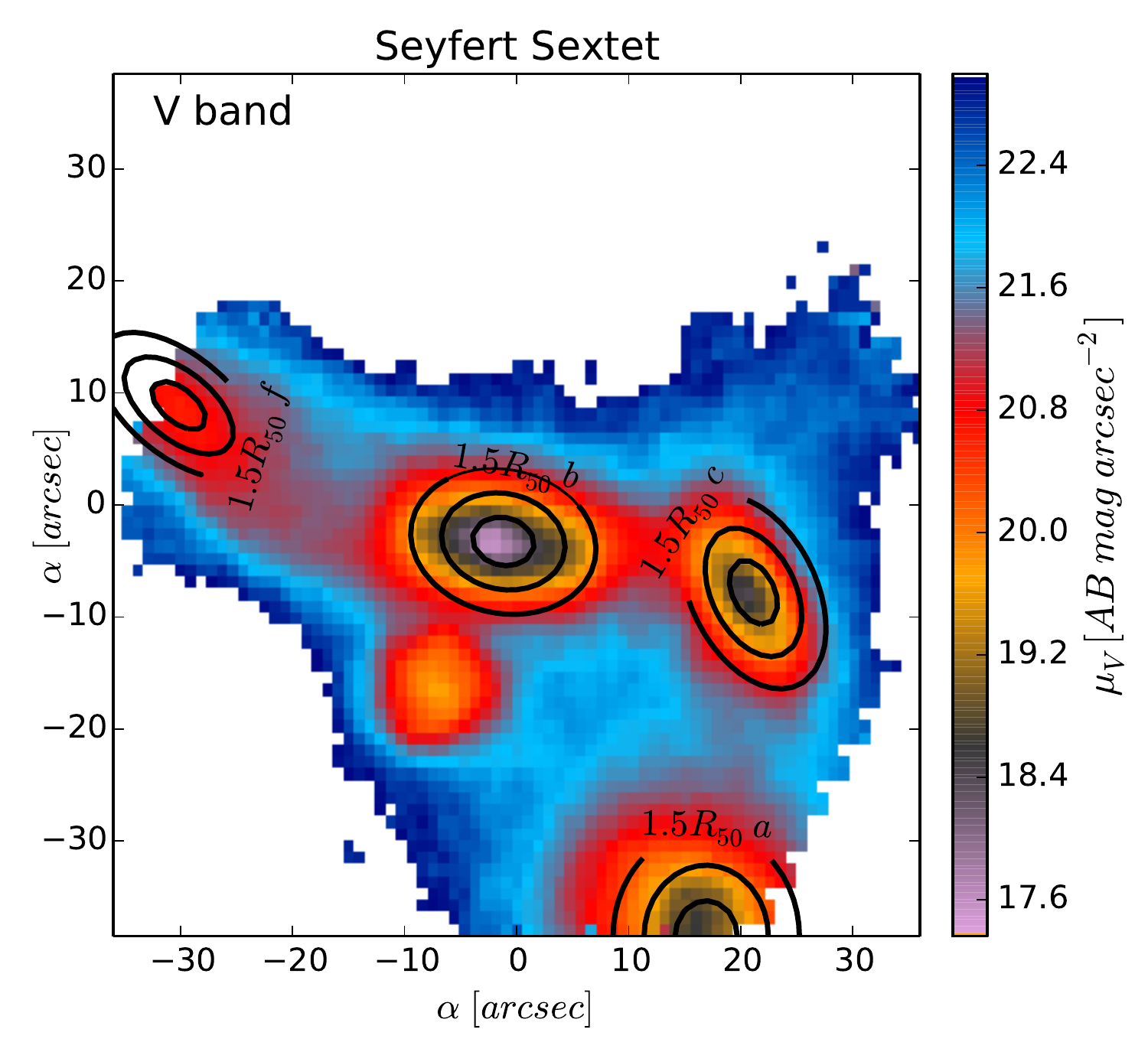}{0.5\textwidth}{(a)}}
 \gridline{\fig{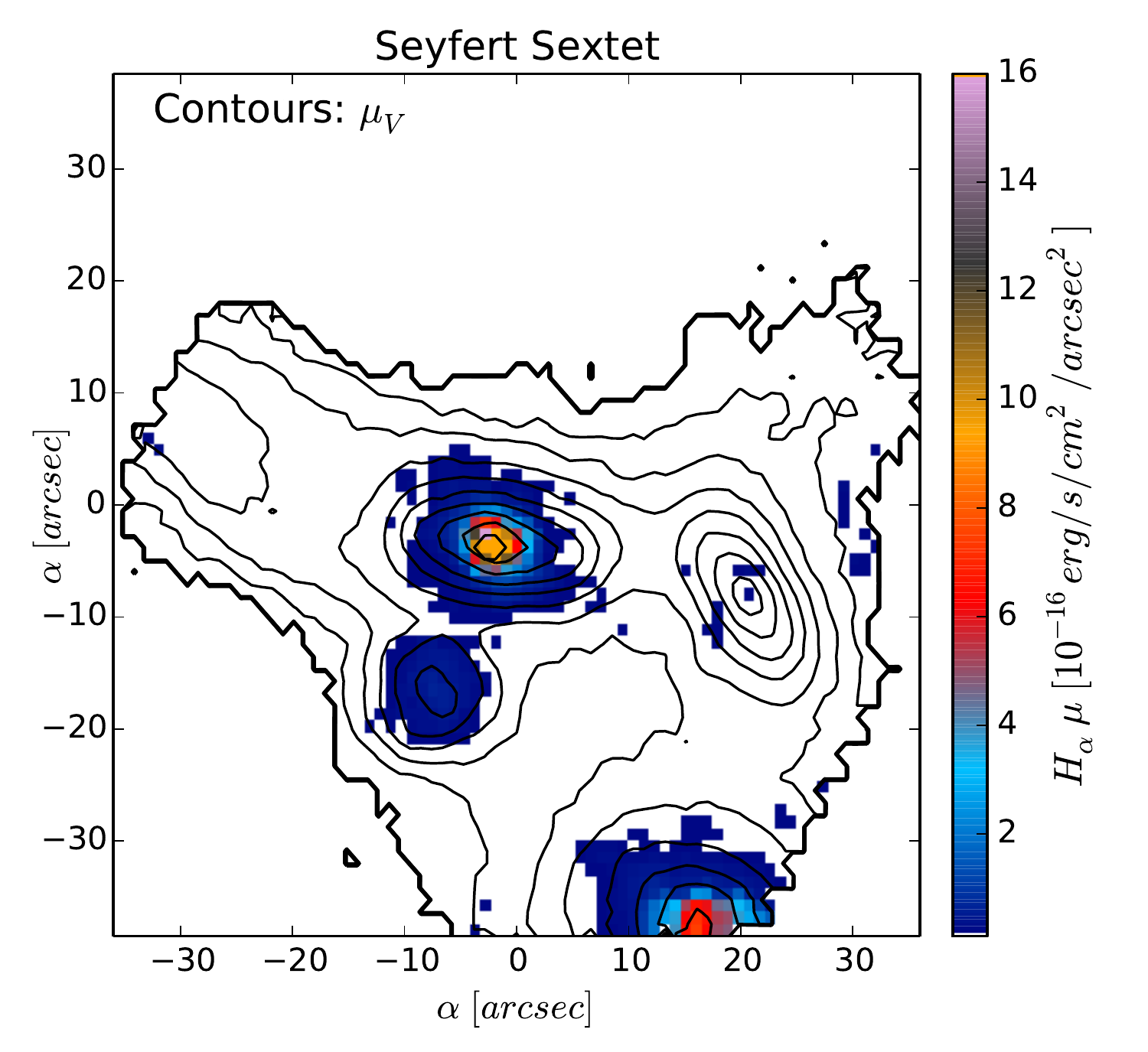}{0.5\textwidth}{(b)}}
 \caption{
  (a) Surface brightness distribution for SS galaxies. The ellipses
  delimit the inner ($0.0R_{e} < R < 0.5R_{e}$), intermdiate 
  ($0.5R_{e}< R < 1.0R_{e}$), and outer ($1.0R_{e} < R < 1.5 R_{e}$)
  regions of interest as multiples of the projected effective radius,
  $R_{e}$. See \S\ref{surface} and \S\ref{starlight} for further
  details.
  (b) $3\arcsec/\mathrm{spaxel}$ resolution map of the
  $\mathrm{H}{\alpha}$ emission overlaid on $V$-band contours
  (black continuous lines) covering the galaxies and intra-group
  light. Coordinates are referred to H79b. The outermost
  contour indicates an isophote with  
  $\mu_{V}=23\; \mathrm{mag\,arcsec^{-2}}$. Contours are spaced in 
  steps of 
  $\mu_{\Delta}=0.6\; \mathrm{mag\,arcsec^{-2}}$. The flux scale is
  indicated in color, above a lower limit of 
  $1.2\times10^{-16}\; \mathrm{erg\,s^{-1}\,cm^{-2}\,arcsec^{-2}}$.
 }  
 \label{fig2}
\end{figure*}

\section{Observations and Analysis Techniques}\label{obs}

\subsection{CALIFA DR3 Observations}\label{califa}
CALIFA DR3 contains IFS datacubes for 939 low-$z$ ($0.005 < z < 0.03$)
galaxies, selected by diameter \citep{Sa16a}. The data were taken with
the Calar Alto Observatory's 3.5-m telescope and the Postdam Multi
Aperture Spectrograph (PMAS) \citep{Ro05}, using the PPAK \citep{Ke06}
integral field spectrophotometer mode. This mode covers a hexagonal
field-of-view (FOV) of $74\arcsec\times 62\arcsec\;(1.3\,\sq \arcmin)$ 
with 331 fiber spectra, each of diameter 
$\diameter_{\mathrm{ fiber}}=2\farcs7$ on the sky, while
the median seeing at Calar Alto is around 1\arcsec.
We report V500 grating observations,
covering $3745\mathrm{\AA}\mathrm{-}7500\mathrm{\AA}$, with spectral
resolution ${\lambda}/{\Delta\lambda} \thicksim 850$, and spatial
resolution $2\farcs5$.  
Three dithered 900-s integrations were taken per target galaxy, using
a pattern of offsets (0,0), (-5.22,-4.53), and (-5.22,4.53) arcsec in
$(x,y)$ relative to the nominal position of the target. This strategy
permits a 100\% filling factor across the field of view. The fluxes
per fiber were initially calibrated by observing standard stars with
some early-type galaxies defined as secondary standards, taking
account of information on observatory sky conditions
\citep[][]{Sa16a,GB15}. The registration procedure is described in
full by \citet{GB15}. The three dithered exposures were combined in a
single frame containing all 993~spectra. Sky-subtrated SDSS DR7 images
were used to recalibrate the fibers using the corresponding fluxes for
331 apertures on the image of the target galaxy. The apertures were
shifted by right ascension and declination inside a search box
centered on the galaxy of interest, with the best registration
determined by minimizing $\chi^2$ for the differences between the
spectra of the raw stack and the SDSS aperture-matched fluxes. An
accuracy of $\sim 0\farcs2$ on the fiber positions was obtained and
the CALIFA spectrophotometry was anchored to SDSS DR7. After Galactic
extinction correction using the \citet{Sch98,Ca89} prescriptions and
correcting for differential atmospheric refraction, 
spectra were rearranged using a flux-conserving inverse-distance
weighting scheme to generate the final data cubes 
\citep[][]{GB15}. Further details of data selection, observational
strategy, data reduction, and data products can be found in
\citet{Sa16a} and CALIFA
DR3\footnote{\url{http://www.caha.es/CALIFA/public_html/?q=content/califa-3rd-data-release}}. 


The observations reported in this paper come from a data cube centered
on H79b (\object{NGC 6027}). However, due the extreme compactness of
SS, four additional galaxies fell within the PPAK FOV. The
$3\sigma$ detection limit per spaxel and spectral resolution element is
$I_3 = 1.3 \times 10^{-18} \; \mathrm{erg\,s^{-1}\,cm^{2}\,\AA^{-1}\,arcsec^{-2}}$
in the median at $5635$ \AA, resulting in a limiting surface
brightness of $\mu_{r}=23.4\; \mathrm{mag\,arcsec^{-2}}$ ($3\sigma$).
The overall spectrophotometric calibration accuracy is better than 
$5 \%$. We used pipeline P\textsc{ipe}3D \citep{Sa16b,Sa16c} for data 
manipulation, masking, spectral extraction, and to estimate the star
formation history (SFH).

Figure \ref{fig1} shows CALIFA's footprint on SS. About half of H79a
is covered by the CALIFA pointing, barely including the nuclear
regions. H79b and H79c are completely covered, while more than three
quarters of H79f are covered. The late-type H79d lies outside this
CALIFA datacube.  Spectra of the galaxies integrated to $1.5R_{e}$
(see \S\ref{surface}) are also shown.  These spectra resemble 
early-type galaxies: the $4000\, \mathrm {\AA}$ break is clear and
the Mgb $\lambda 5175$ and NaD $\lambda5892$ absorption lines are also
seen. Emission in $\mathrm{H}{\alpha}$ is seen in H79a and H79b, and
in $\mathrm{H}{\beta}$ in H79c.  Fig~\ref{fig2}(b) shows emission maps
of $\mathrm{H}{\alpha}$ for H79a, H79b and H79c.  

\begin{deluxetable}{lccccccccc}
 \tabletypesize{\scriptsize}
 \tablecaption{Surface Brightness Parameter for Seyfert's Sextet Galaxies \label{tbl-1}}
 \tablewidth{0pt}
 \tablehead{
  \colhead{Galaxy} &\colhead{Model} & \colhead{$\mathrm{r_{e}}$} &
  \colhead{$R_{e}$} &\colhead{$n$} &\colhead{$e$} & \colhead{P.A.} &
  \colhead{$m$} & \colhead{B/T} & \colhead{$m_{T}$}\\ 
  &&\colhead{[\arcsec]}&\colhead{[kpc]}&&&[\arcdeg]&[mag.]&&[mag.]
 }
 \startdata
 H79a&{S\'ersic}&$8.3\pm0.7$&$2.3\pm0.2$&$3.5\pm0.2$&$0.325\pm0.005$&$72.0\pm0.5$&$13.81\pm0.06$
 && \\ 
 -&Disk\tablenotemark{a}&\nodata&&\nodata&\nodata&\nodata&15.7&0.85&13.63\\
 &&&&&&&&&\\
 H79b
 &{S\'ersic}&$6.7\pm0.6$&$1.8\pm0.2$&$1.6\pm0.3$&$0.73\pm0.00$&$82.13\pm0.08$&$14.2\pm0.1$&&\\ 
 &Exp. Disk&$0.75\pm0.06$\tablenotemark{b}&$0.21\pm0.02$&\nodata&$0.52\pm0.05$&$74.00\pm0.03$
  &$16.8\pm0.2$&0.92&14.12\\ 
 &&&&&&&&&\\
 H79c
 (bulge)&{S\'ersic}&$0.24\pm0.02$&$0.07\pm0.01$&$1.8\pm0.1$&$0.14\pm0.01$&$34\pm5$&$20.41\pm0.08$
 && \\ 
 - (bar)
 &{S\'ersic}&$1.05\pm0.01$&$0.290\pm0.003$&$0.5\pm0.0$&$0.545\pm0.005$&$33\pm1$&$18.32\pm0.03$
 && \\ 
 -&Exp. Disk&$5.67\pm0.07$\tablenotemark{b}&$1.56\pm0.02$&\nodata&$0.61\pm0.01$&$35\pm1$
 &$15.37\pm0.01$&0.07&15.29\\ 
 &&&&&&&&&\\
 H79f&{S\'ersic}&$9.1\pm1.1$&$2.5\pm0.3$&$0.53\pm0.01$&$0.47\pm0.01$&$57\pm5$&$16.0\pm0.1$ 
 &\nodata&16.0\\
 \enddata
 \tablenotetext{~}{COLUMNS-- 1: name of the galaxy,  2: model fit, 3:
  effective radius ({\em arcsec}),4: effective radius (kpc), 5:
  S\'ersic index, 6: ellipticity, 7: position angle ({\em deg}),
  8: integrated magnitude of each component, 9: bulge-to-total ratio,
  10: apparent total magnitude} 
 \tablenotetext{a}{No fit for the disk component was possible, light
   was integrated on the residual image.} 
 \tablenotetext{b}{ For exponential disk fits, we report
   $\mathrm{r_{e}=1.678r_{h}}$, where $\mathrm{r_{h}}$ is the disk
   scale length.} 
\end{deluxetable}

\subsection{2D Surface Brightness}\label{surface}
A flux-calibrated, $2000\,\mathrm{s}$ integration image taken with
HST/WFPC2 ($0\farcs1\,\mathrm{pixel^{-1}}$) in the
F814W\footnote{Hubble Legacy Archive, which is a collaboration between 
 STScI/NASA, ST-ECF/ESA, and CADC/NRC/CSA, \url{https://hla.stsci.edu}
 and  \url{http://www.stsci.edu/hst/wfpc2}} band was used. We use the
2-D parametric code GALFIT \citep{Pe10} to model the surface
brightness distribution of SS galaxies. We have used the sum of one or
two S\'ersic models, plus an exponential disk to account for bulge,
bar, and disk components, respectively. A point spread function
(PSF) generated by Tiny
Tim\footnote{\url{http://www.stsci.edu/software/tinytim/}} was
employed in this study. GALFIT convolves the analytical models (a
surface of a single orientation, which defines the position angle)
with the PSF and performs a $\chi^2$ minimization between the
convolved model and pixels on the galaxy image. See the
GALFIT\footnote
{\url{https://users.obs.carnegiescience.edu/peng/work/galfit/galfit.html}}
home page for further details.  WFPC2 is a four CCD mosaic; to cover
the gaps between CCDs, multiple shifted images are taken and then
combined.  Four 500 sec images were
DRIZZLEd\footnote{\url{http://www.stsci.edu/scientific-community/software/drizzlepac.html}}
to generate the image that we used, but some gaps with small negative
counts remained. Gaps and galaxy dust lanes were masked.    

Diffuse intra-group light \cite[e.g.,][Du08]{DR05}, 
crowding, and tidally-induced asymmetries complicate the
modeling of the surface brightness.  Hence we proceeded interactively,
galaxy by galaxy, considering regions of increasing sizes. In
a second pass, all the galaxies were modeled simultaneously, using the
previous fits as inputs.  The entries in Table~\ref{tbl-1} are average
parameters, the errors represent the dispersion for each parameter.
Using the largest effective radius, $R_{e}$, of the fitted components
for each galaxy, we consider the following regions:  
inner ($0.0R_{e} < R < 0.5R_{e}$), 
intermediate ($0.5R_{e}< R < 1.0R_{e}$), and 
outer ($1.0R_{e} < R < 1.5 R_{e}$), see Fig.~\ref{fig2}(a). 
These regions will be used in further analysis  below. 

\subsection{Fossil Stellar Record}\label{starlight}

Among the many packages and pipelines that are available to recover
the fossil stellar record from the  spectral energy distributions
(SED) of
galaxies\footnote{\url{http://www.sedfitting.org/Fitting.html}}
\citep[e.g., reviews by][]{Wa11,Co13}, we chose
P\textsc{ipe}3D\footnote{\url{http://www.astroscu.unam.mx/~sfsanchez/FIT3D/}}
\citep{Sa16b,Sa16c}. This pipeline performs a set of complex
operations  on  the input  data, and uses on the inversion method.
The steps within P\textsc{ipe}3D are broadly described  below: for
detail the reader is referred to \citep{Sa16b,Sa16c, Ib16, Sa19, Ib19}.

The first function of P\textsc{ipe}3D applies spatial binning, when
necessary, to reach the minimum signal-to-noise ratio (S/N) needed to
extract a realistic decomposition of the stellar population 
\citep{Sa16b}.  Then corrections are applied to account for stellar
kinematics and dust attenuation for each (possibly spatially binned)
spectrum. A limited single-stellar population (SSP) template
library is used to limit and partially avoid the degeneracies from
dust, metallicity, age, and velocity dispersion. This fitting 
involves a non-linear minimization (for dust and kinematics
properties), and a liner fit for the combination of the dust-obscured,
velocity shifted and broadened version of the SSP templates. The
kinematics and stellar dust attenuation parameters are then fixed for
the rest of the analysis. Linear fitting is repeated using an  
extended SSP-library\footnote{The full SSP library is held at
\url{https://svn.sdss.org/public/data/manga/pipe3d/trunk/data/BASE.gsd01}}.
During  a first iteration the emission line regions are masked, to
minimize their potential effect on the fit. A first model for the
stellar population is then generated and removed from the original
data to create a ``pure gas" (plus noise) spectrum. The emission lines
are then fitted by a set of Gaussians, and the resulting emission
model is removed from the original spectrum, creating a ``pure
stellar-population" spectrum that it is then fitted again with the
SSP-library. The process is repeated until convergence is reached
(normally after three iterations).

Errors for each fitted parameter are estimated using Monte Carlo (MC)
simulations of the original spectrum. These take into account dust
attenuation of the stellar population, the velocity and velocity
dispersion, the flux intensity, the velocity and velocity dispersion
of the considered emission lines, and the weights of the
decomposition of the stellar population in the SSP library. The
process is repeated in its entirety for 30 times.  Using the SSP
library weights, we derive the Mass-Assembly History (MAH) following
the procedure described in \citet{Sa16c, Sa19}. First, we multiply the
weight of each SSP, normalized to at a predefined wavelength, by the
flux intensity at that wavelength, and the Mass-to-Light ratio of that
SSP. This give us the distribution of masses at each age and
metallicity. Then we sum the masses at each age to create the
cumulative function 
\citep[correcting for the mass loss, following][]{Co14}, i.e., 
the MAH. We assume that mass variations are due to star-formation.  
Hence, by calculating the  differential mass at a given epoch (defined
by the age sampling in the SSPs) and dividing by the time interval, we
obtain the star-formation rate (SFR) at each epoch \citep{Sa19,Ib19}.
Using this information, we can generate the star formation history
(SFH).  Dividing the SFR by the assembled mass at each time, we obtain
the specific star-formation history (sSFH). This definition is
employed  throughout  this paper. Errors are propagated from the MC
process described above.  

Comparisons with other codes and simulations have been used to
validate P\textsc{ipe}3D, as detailed by 
\citet{Sa16b, Ib16, Sa19, Ib19}. For S/N $> 50$, $A_{V}$ is recovered
with an accuracy of about 0.06~mag and a precision of about 0.17~mag.
A more important error arises from galaxy inclination
\citep{Ib19}. In our case, due to the distorted morphologies of the SS 
galaxies, it is difficult to assess the inclination realistically, so
that it is difficult to estimate the magnitude of the consequent
error. On the other hand, the metallicity evolution of galaxies has
not been checked in detail within P\textsc{ipe}3D. Nevertheless,
\citet{Sa16b} showed that P\textsc{ipe}3D closely matches the
results of \textsc{starlight} \citep{CF05}. We therefore expect
that P\textsc{ipe}3D can reproduce \textsc{starlight}'s chemical
evolution results. A general pattern of stellar populations having
lower metallicity in the past than the present is expected to emerge,
but we also expect the stellar metallicity to be closely related to
the stellar mass in spheroidal galaxies, but rather to relate to
stellar surface mass density in disk-dominated galaxies
\citep[e.g.,][]{GD14, CF15}.

We have evaluated the SFHs per spaxel using P\textsc{ipe}3D.  Our
template library contains 156~spectra to cover 39 stellar ages, from
1~Myr to 14.2~Gyr, and four metallicities ($Z/Z_\sun$ = 0.2, 0.4, 1.0
and 1.5). P\textsc{ipe}3D processed galaxies from SS and the control
sample via brute-force fitting, exploring in fine steps all possible
linear combinations of the library spectra, to obtain the best fit
with no assumptions on the shape or functional form of the SFHs. 
The radial structure of the sSFHs \citep[e.g.,][]{Pe13} is generated
by an implementation by \citet{Ib16} based on inner, intermediate and
outer regions as shown in Fig.\ref{fig1}(b) (see \S\ref{surface} and
Table~\ref{tbl-1}). On average, the inner regions of galaxies H79a,
H79b, and H79c have $S/N \thicksim 200$, and the outer regions have
$S/N\thicksim 50$. H79f has $S/N \thicksim 50-60$ across the region
covered.

\subsection{Kinematics as Traced by Stars and Ionized Gas in Galaxy H79b}\label{dynamics}
The data for H79b have high S/N, and H79b is completely covered by the
CALIFA footprint. A separate analysis was performed to retrive kinematic
information following the technique described by \citet{Ba14,GL15}. On
the first pass, spaxels are Voronoi-binned, ignoring spaxels with
$\mathrm{S/N<3}$, producing voxels, discrete regions where
$\mathrm{S/N\geq20}$. To derive the line-of-sight velocity maps, each
voxel is fitted, avoiding emission lines, using  pPXF \citep{CE04}.
Errors per spaxel are estimated via Monte Carlo simulations and range from
$5\mathrm{-}20\,\mathrm{km\,s^{-1}}$ \citep{Ba14}. Ionized gas
maps are generated after the stellar contribution is removed. The
velocities are derived using the cross-correlation method implemented
by \citet{GL15}.  Each spectrum (whose $\mathrm{S/N>8}$) is compared
with a template that includes the 
$\mathrm{H_{\alpha}+[NII]\,\lambda\lambda 6548,6684}$  emission lines
in the interval
$6508\,\mathrm{\AA}\mathrm{-}6623\,\mathrm{\AA}$. Templates are 
convolved with Gaussians to account for the instrumental resolution
and shifted to the galaxy systematic velocity. Typical errors in the
velocities are of order of $10\, \mathrm{km\,s^{-1}}$. Stellar  
and ionized-gas velocity maps are shown in Figure~\ref{fig5}(a).

\subsection{Dust Properties}\label{dust}
We used an independent indicator of star formation to check the
results from \S\ref{starlight}. Although a more detailed analysis has
been provided by \citet[][]{Bi11, Bi14}, we wanted to compare  
the results from a simple but robust method using additional data. 
We applied the code {\tt cmcirsed} developed by \citet{Ca12} to model
the IR spectral energy distribution (SED) in wavelength range 
$8 - 500\,\mu\mathrm{m}$, generated from Spitzer, WISE, and Herschel 
data. {\tt cmcirsed} fits these data with a modified
single-temperature grey body plus a power law. We performed the fit 
for H79a, H79b and H79c.  After allowing the emissivity index ($\beta$)
and the mid-IR power-law slope ($\alpha$) to vary, we derived the peak
wavelength dust temperature $T_{\mathrm{dust}}$, the infrared luminosity
integrated from $8\,\mu$m up to $1000\, \mu$m, $L_{\mathrm{IR}}$, and
the dust mass 
$M_{dust} \propto D_{\mathrm{L}}^2/[ \kappa _{\nu}\, \nu^2 (1+z)]$, 
where $D_{\mathrm{L}}$ is the luminosity distance, $\kappa _{\nu}$ is
the dust mass absorption coefficient at frequency $\nu$ 
\citep[see][for further details]{Ca12}. 
The fits are presented in Table~\ref{tbl-2} \citep[cf.,][]{Bi11,Bi14}.

\begin{deluxetable}{lccccc}
 \tabletypesize{\small}
 \tablecaption{Infrared Properties for Galaxies in the Seyfert's
  Sextet modeled using {\tt cmcirsed} \citep{Ca12} \label{tbl-2}} 
 \tablewidth{0pt}
 \tablehead{
  \colhead{Galaxy} &\colhead{$\beta$} & \colhead{$\alpha$} &
   \colhead{$\log L_{\mathrm{IR}}$} &\colhead{$\log M_{\mathrm{dust}}$}
   & \colhead{$T_{\mathrm{dust}}$} \\ 
  &&&\colhead{[$L_{\sun}$]}&\colhead{[$M_{\sun}$]}&[K]
 }
 \startdata
 H79a&$2.0\pm0.0$&$1.61\pm0.04$&$9.57\pm0.04$&$6.59\pm0.06$&$22\pm 1$\\
 H79b&$2.3\pm0.6$&$2.0 \pm0.1 $&$9.5 \pm0.1 $&$5.1 \pm0.1 $&$25\pm 2$\\
 H79c&$0.6\pm0.4$&$3. 6\pm0.3 $&$8.6 \pm0.1 $&$5.1 \pm0.1 $&$28\pm 3$\\
 \enddata
 \tablenotetext{~}{COLUMNS-- 1: name of the galaxy,  2: Emissivity, 3:
  Mid-IR power-law slope, 4: infrared  Luminosity ($L_{\sun}$), 5:
  Dust Mass ($M_{\sun}$), 6: dust temperature (K)} 
\end{deluxetable}

\section{A Framework for Group's Dynamics}\label{model}
\citet{Be85} considered self-similar secondary infall and accretion,
in an  Einstein-de Sitter Universe, as an improvement over spherical
gravitational collapse \citep[e.g.,][]{vdWB08}. In this model
overdensities decouple from the cosmic expansion at some maximum
radius, the first turn-around radius $r_{1t}$, and then
collapse. After crossing, overdensities rebound, but with smaller
amplitude, turn around and recollapse 
\citep[cf., CG numerical simulations by][]{Ba85}. The
pause before the recollapse generates discontinuities in the density
and velocity dispersion as functions of radius. 
\citet[][hereafter Tu15]{Tu14} suggested that such discontinuities are
observable in the caustic that defines the second turn-around radius
$r_{2t}$. \citeauthor{Tu14}'s insight led to a tight
correlation between a group's line-of-sight velocity dispersion,
$\sigma_p$, and the projected second turn-around radius, $R_{2t}$, 
\begin{equation}
 \frac{\sigma_{p}}{R_{2t}} = (368\pm 8)\,h_{75} 
  \,\mathrm{km\,s^{-1}\,Mpc^{-1}},
  \label{eqn-1}
\end{equation}
valid for $70 \le \sigma_{p}/\mathrm{km\,s^{-1}} \le 1000$, i.e., from
galaxies to galaxy clusters,
where $\sigma_{p}=({\Sigma_{i}(v_{i}-\bar{v})^2}/{n})^{\onehalf}$ and
$\bar{v}$ is the group's mean radial velocity, for  $i=1...n$ group
members with radial velocities $v_i$.
For spherical symmetry the second turn-around radius is related to the
projected turn-around radius $r_{2t}=\sqrt{{3}/{2}}\,R_{2t}$.  A
direct application of the virial theorem leads to 
$R_{2t} = 0.215 \, M^{1/3}_{12}\, h^{-2/3}_{75}\; \mathrm{Mpc}$, 
where the virial mass
$M_{v} = 2\times 10^{6} \, ({\sigma_{p}}/{\mathrm{km\, s^{-1}}})^{3}\,h^{-1}_{75}\;M_{\sun}
       = M_{12}\times 10^{12}\;M_{\sun}$  
(cf., Du08, \S3). Estimating the first turn-around radius, $r_{1t}$,
requires 3D information to distinguish the decoupling of the infall 
zone from the Hubble flow. Such information is currently very limited.
Nevertheless, Tu15 used direct measurements for Virgo, the M~82 group, 
and the local association (see below), and found an estimate of
$r_{1t}=0.77\,M_{12}^{1/3}h^{-2/3}_{75}\; \mathrm{Mpc}$. Since
both turn-round radii should enclose the same mass, we have 
${r_{1t}/r_{2t}} = 3.14\;\pm\, 0.28$. This ratio is partially
justified by theory, see Tu15 for further details. 
 

Using the velocities for H79a, H79b, H79c, H79d and H79f taken from Du08 
and our own measurement using CALIFA data, we find
$\bar{v}=4296 \; {\rm km\, s^{-1}}$ and
$\sigma_{p}= 160\,\pm\,72\; {\rm km\, s^{-1}}$. From
Eq.~\ref{eqn-1} we then find 
$R_{2t}= 435\,h_{75}^{-1}\;\mathrm{kpc}$,
and hence 
$r_{2t}\approx 533 \,h_{75}^{-1}\;\mathrm{kpc}$, 
$r_{1t}\approx 1673 \,h_{75}^{-1}\;\mathrm{kpc}$, and
$M_{v}\approx8.2\times10^{12}\;h^{-1}_{75}\;M_{\sun}$, 
i.e., $M_{12}=8.2$.  

Tu15 also introduced {\em associations} as quasi-virialized regions
where giant galaxies along with their respective satellite galaxies
are identified as groups. For example, the Milky Way and M~31 are
defined as two individual groups within the {\em local association}
\citep[cf.,][]{LB94}. The dynamical history of these galaxy-scale
groups could be as complex as in CG --- for example, it has been
suggested that the dwarf satellite galaxy M~32 originates from
the disruption of a galaxy comparable to the Milky Way due to an
interaction with M~31 \citep{DS18}.

The \citeauthor{Be85} secondary infall scenario applies to any
overdensity, hence it applies to galaxy-scale groups such as the
Milky Way and M~31, and to any isolated galaxy halo. In these 
cases the turn-around radii can be calculated by Equation~\ref{eqn-1} 
and the formalism above applies directly. This can be recognized as
self-regulated galaxy evolution \citep[e.g.,][]{TM10}, as
well. Nevertheless, within the current cold dark matter paradigm other
effects may be at play, for example AGN feedback
\citep[e.g.,][]{SCD14}.

\citet{LB81,LB94,LB99} developed a simple dynamical model for groups
of galaxies, in which the solution for the equation of motion can be
approximated by the following polynomial \citep[Eq. 13 in][]{LB94}: 
\begin{equation}\label{approx}
 \left(\frac{GM_{gp}}{r^3}\right)^{\onehalf}t+0.85
 \left(\frac{v}{r}\right)t=2^{-\frac{3}{2}}\pi \approx 1.11, 
\end{equation} 
where $G$ is the gravitational constant, $M_{gp}$ is the total group
mass, $t$ is the time, $r$ is the separation between galaxies, and
$v$ the velocity of the galaxies.  At the turn-around radius, $r_t$,
the group galaxies are neither approaching nor receding from one
another, and so the time of turn-around is given by
\begin{equation}\label{turnaround}
t=1.11 \left({\frac{r_t^3}{GM_{gp}}}\right)^{\onehalf},
\end{equation} 
which is also approximately the free-fall time. For a
given radius, more massive groups will turn around faster.

If we adopt $M_{gp}=10\times 10^{12}\, M_{\sun}$; then 
$t_{1t} \sim 11$~Gyr and $t_{2t} \sim 2$~Gyr. 
Therefore, SS is able to cross at least once within 
$t_H$. We suggest that by using the fossil record of star formation of
CG galaxies, we can identify  the corresponding  crossings as
tidally-induced coeval bursts of star formation. This gives another
timing argument to help us constrain the  evolution of CG (see
\S\ref{sSFH}). 


\section{Discussion}\label{dis}

\begin{figure}[t]
 \includegraphics[scale=.70]{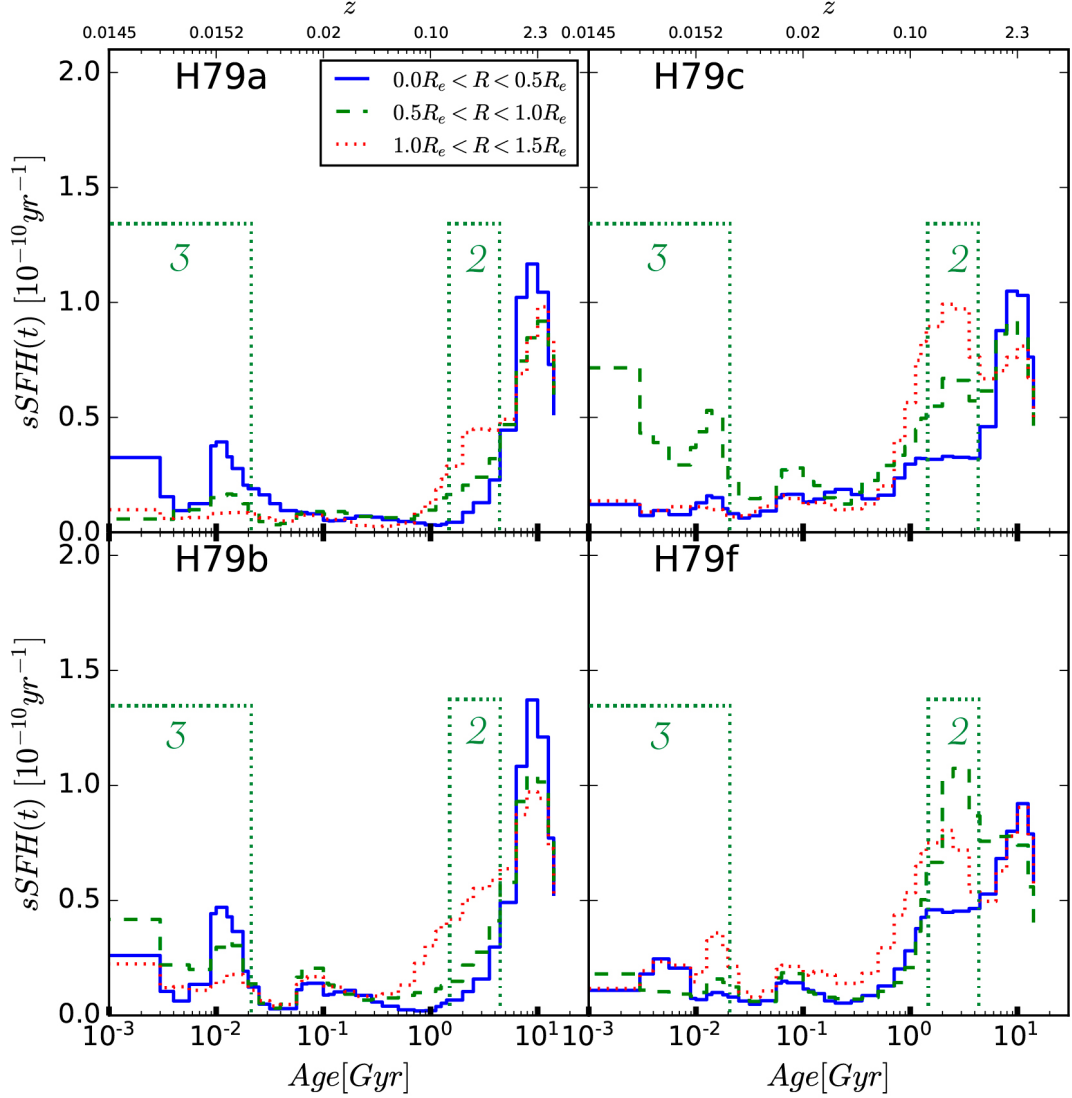}
 \caption{sSFH for SS galaxies. The age of the stellar
  population, given in Gyr, is indicated on the lower $x$-axis. The
  upper $x$-axis indicates the redshift $z$, as a proxy for the
  look-back time. The strength of the sSFH is indicated on the
  $y$-axis, expressed  in units of $[10^{-10}\, yr^{-1}]$. The lines
  indicate the inner (blue), intermediate (green), and outer (red)
  regions of the galaxies.  The green rectangles labeled 2 and 3 mark
  coeval episodes of enhanced star-formation.
 }  
 \label{fig3}
\end{figure}

\begin{figure*}[t]
 \gridline{\rotatefig{270}{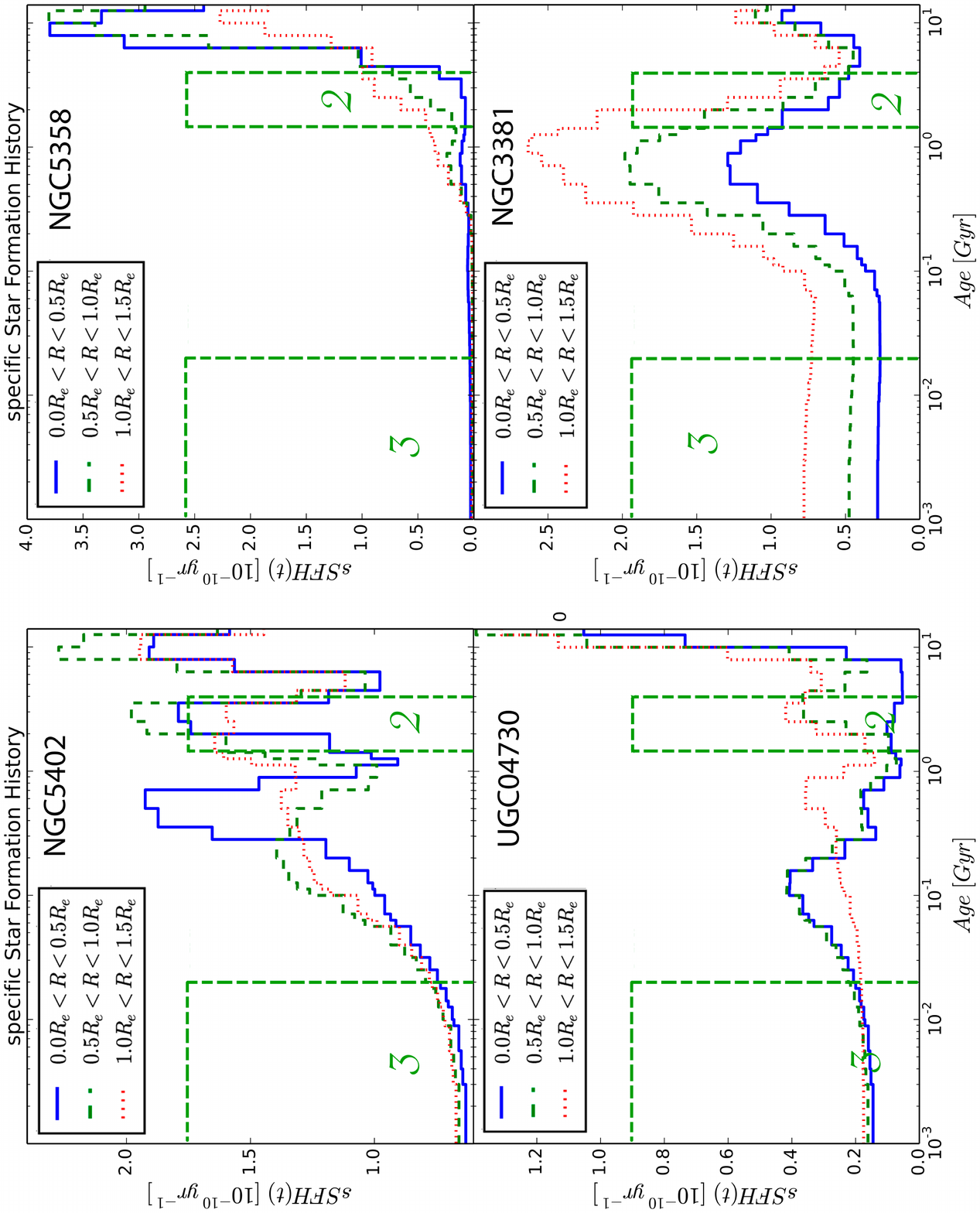}{0.6\textwidth}{(a)}}
 \gridline{\rotatefig{270}{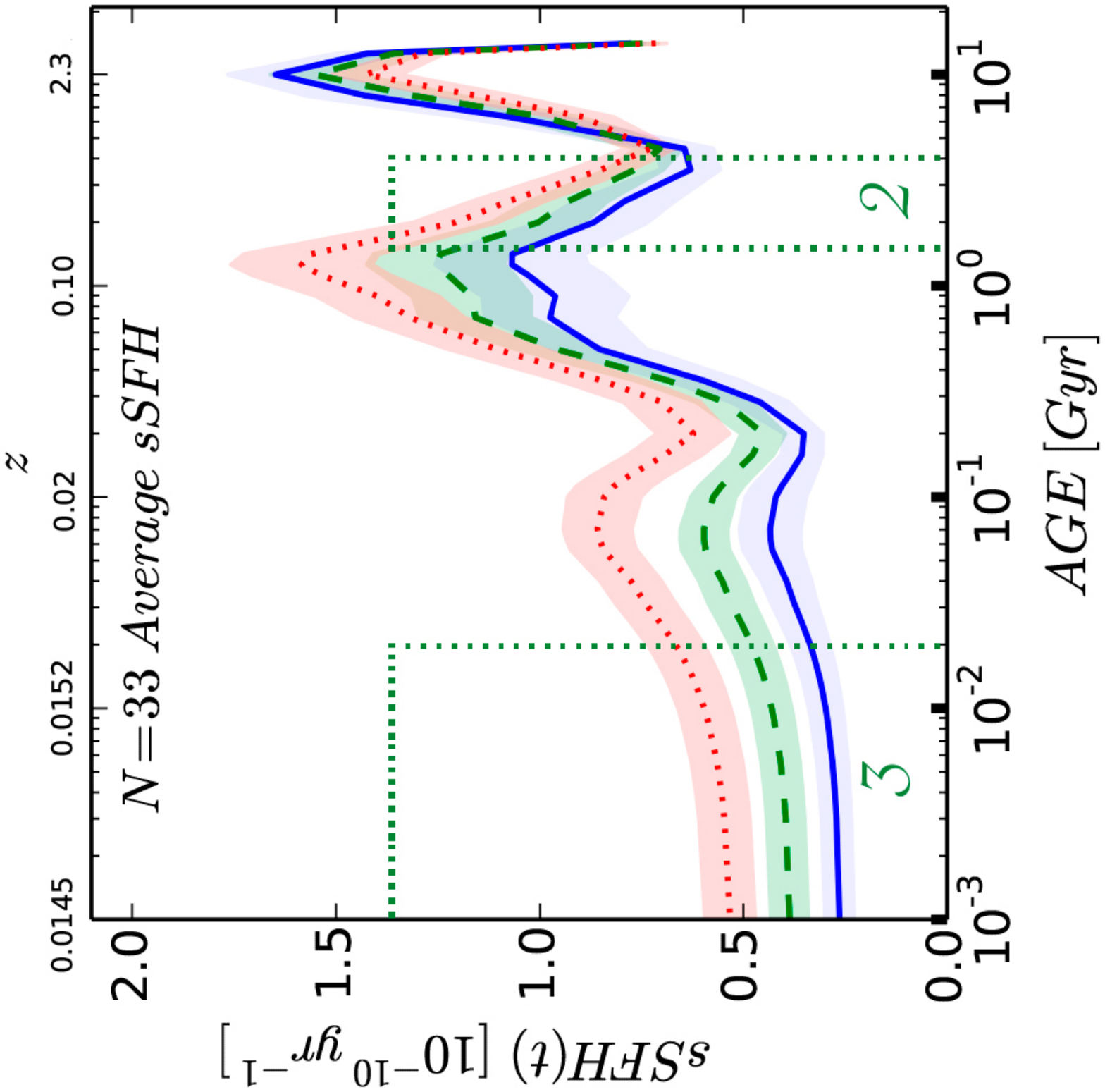}{0.7\textwidth}{(b)}}
 \caption{
  (a) sSFH for the control galaxies NGC~3381, NGC~5358, NGC~5402 and
   UGC~4730. 
  (b) The average sSFH for 33 control galaxies of similar mass and
   redshift as SS galaxies. All data have been taken from the CALIFA
   sample. The notation is the same as in Figure \ref{fig3}.
 }  
 \label{fig4}
\end{figure*}

\subsection{Star Formation History (SFH)}\label{sSFH}
Great progress in modeling stellar populations has been
achieved by the use of extensive stellar libraries,
better optimization techniques, and reliable error analysis. It
is now possible to trace the evolution of individual galaxies more
accurately \citep[e.g.,][]{Wa11,Co13, Ib16,Sa19, Ib19}.  Hence, the
analysis of the sSFH could, in principle, provide us with timing
arguments. Hereafter, we assume that group galaxies undergo
simultaneous bursts of star formation induced by tidal interaction at
each group crossing.

We have restricted our analysis to mass-weighted stellar
age. Figure~\ref{fig3} shows the sSFH for H79a, H79b, H79c and H79f as
a function of radius. A smoothing of 0.1 dex was applied to account
for the  resolution  of the stellar libraries. 

Figure~\ref{fig3} indicates that the sSFHs of SS galaxies show bursts
of star formation at comparable ages but in different regions inside
the galaxies. Figure~\ref{fig3} shows that H79f's sSFH is as
complex as the other SS galaxies; hence, contrary to \citet{Ni02}, 
we suggest that H79f is a disrupted galaxy rather than a tidal tail
(cf., Du08): H79f's light distribution and low S\'ersic
index $n$ (see \ref{surface} and Table \ref{tbl-1}) suggests that
H79f was originally a late-type galaxy (cf., Du08). 

A closer examination of Figure~\ref{fig3} suggests that the bursts of
star formation in SS galaxies are almost synchronized, with active
intervals marked by labeled green rectangles~2 and~3 in the Figure. We
can test the reality of this synchronization by looking at galaxies
not in SS,  since we would not expect enhanced SF in the same time
intervals.
From the CALIFA DR3 data base we selected a set of galaxies to
serve as a control sample. Since galaxy mass and environment are
important drivers of galaxy evolution, we chose galaxies with masses
comparable to SS galaxies ($M\sim10^{10}M_{\sun}$, see 
\S\ref{rotation}) found in groups, pairs, or the field. We
also demanded that the selected galaxies should be at redshifts
comparable to SS ($z = 0.0145$), to match the age of the SS galaxies
and make comparisons easier. The final control sample has 33
galaxies.  We found a different sSFH pattern (Fig.~\ref{fig4}), with
no coordinated burst of star formation between any pair of galaxies at
any epoch.

Figure~\ref{fig4}(a) shows the individual sSFH for control galaxies
NGC~3381, NGC~5358, NGC~5402 and UGC~4730. Their histories are
representative of typical sSFH patterns that can be found in galaxies
of similar mass. Variations in sSFH may be related to the effects of
mergers or the assembly of their dark matter halos. We can
identify up to three peaks of star formation of varying intensity at
different epochs in Figure~\ref{fig4}(a).  Nevertheless, when we
combine the 33 control galaxies, the peaks are smoothed and the
pattern of self-regulated star formation emerges, as is apparent in 
Figure~\ref{fig4}(b). 


We relate redshift to stellar age ($t_{sa}$, with
$t_{sa}=0$ in the SS rest-frame, $z_{SS}=0.0145$), as shown on the upper
$x$-axis of Fig. \ref{fig3}.  We can identify an
initial starburst at $t_{sa} \gtrsim 10$ Gyrs ($z \gtrsim 2.3$), which
is also seen in the control sample (Figs. \ref{fig4}(a,b)), we label
this starburst~1, which is likely of cosmological origin.

Assuming that tidally-triggered star formation acts instantaneously,
we identify the sSFH peak at stellar age 
$t_{sa} \thicksim 3$~Gyr ($z_{fc} \thicksim 0.3$) with the first
crossing, after a first turn-around at $t_{1t} \sim 11$~Gyr.
Group-wide interactions at $t_{sa}$ triggered bursts of star
formation. This is suggested by the rise of activity in the
intermediate and outer regions of each galaxy. H79a and H79b, being
more massive, evolve faster, but clear extended star formation is seen
in their intermediate and outer regions (green rectangle~2 in 
Fig.~\ref{fig3}). This episode of merger-triggered star formation
lasted for up to about~3.5 Gyr. Most of the original gas 
reservoir would have been used during this epoch, and dust would
have been largely depleted by starburst-induced winds. Hence any later
merger would be expected to be dry or mixed \citep[cf.,][]{Bi11}.

A third peak of star formation is expected after the second
turn-around at $t_{2t}\thicksim2$ Gyr \citep[cf.,][]{PF12}. There are
signs of star formation in the inner regions of H79a and H79b and the
intermediate regions of H79c and H79f that peak around
$t_{sa}=0.02$ Gyr (Fig.~\ref{fig3}), hence we suggest that SS galaxies
began their second crossing at $z_{sc}\thicksim 0.016$ 
(green rectangle~3 on the figure). 
Since a modest level of star formation still continues at
$t_{sa} = 0$, we conclude that we are observing SS during its second crossing.
This may explain why SS is one of the densest galaxy aggregates
known (five galaxies within $37\,h_{75}^{-1} \; \mathrm{kpc}$, cf., Du08). 

\begin{figure*}[t]
 \gridline{\rotatefig{90}{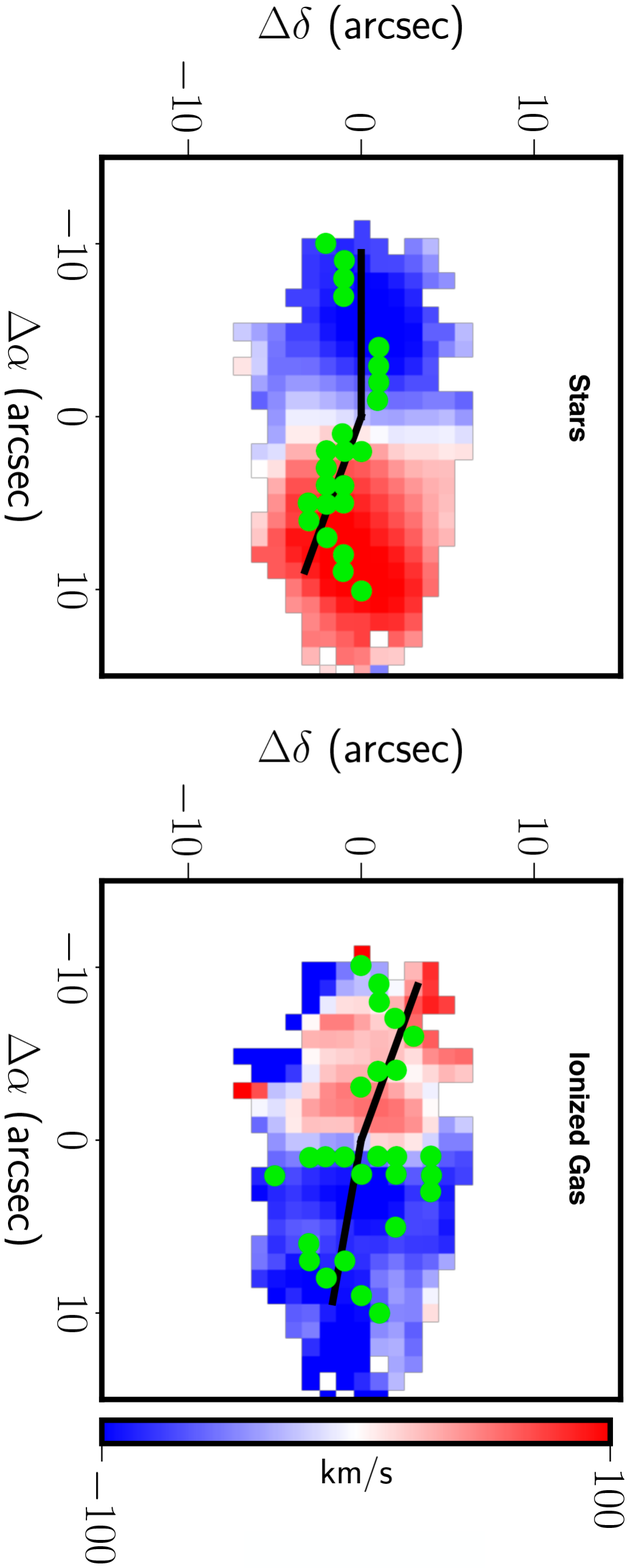}{0.4\textwidth}{(a)}
          \fig{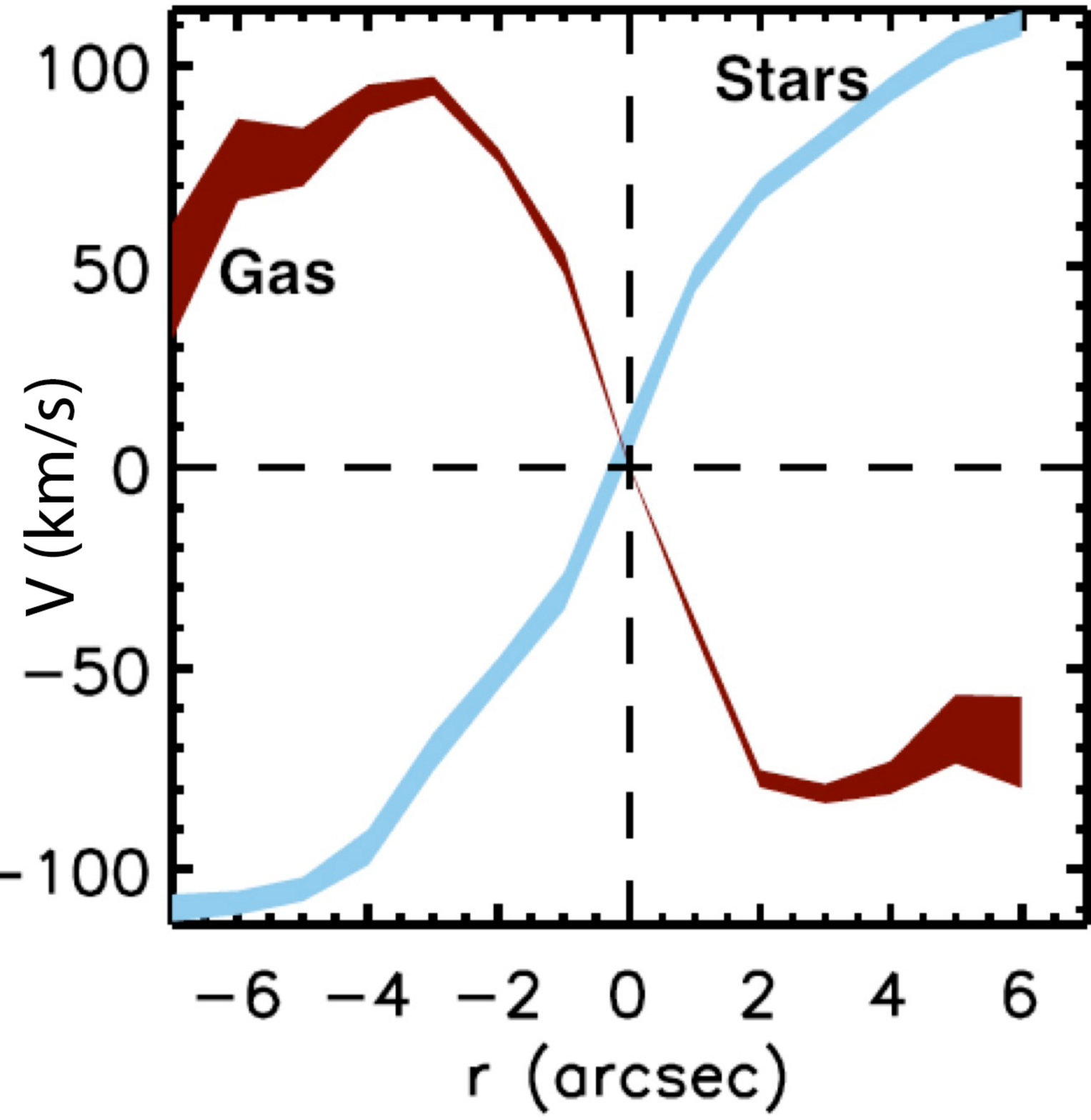}{0.29\textwidth}{(b)}
           }
 \caption{
  A counter-rotating disk in H79b. 
  (a) The velocity fields due to stars and ionized gas,
  respectively. In each panel, green points highlight the radii
  of maximum velocities, determined from the pseudo-rotation curves
  (pRC) of H79b. Black lines represent the average kinematic
  PA. The color-coded velocity scale is shown on the right.  
  (b) The pRCs for the stars (blue) and ionized gas (brown).
 }  
 \label{fig5}
\end{figure*}

The locations of the continuing star formation differ. H79a and H79b
show mild star formation in their inner regions, while in H79c only
the intermediate region is still forming stars. H79f has lost most of
its gas so that only a low level of star formation can be sustained.
Figures~\ref{fig4} and~\ref{fig5} show that SS follows a
different pattern of star formation than the control galaxies. The
second burst of star formation happened slightly earlier than in the
control sample on average. However, the most noticeable  difference
is seen at redshifts around $z = 0.0145$, where the control galaxies
show a monotonic decline of their mean sSFH.

To complete our scenario, we propose that during each crossing some
galaxies might be disrupted (e.g., H79f), losing their stars 
and gas to enrich the intragroup medium \citep[cf.,][]{DR05,Zem14,BYV10}.  

Since SS galaxies have appreciable sSFR at present (Fig.~\ref{fig3},
region 3) we should be able to detect the spectral signature of star
formation. Figure~\ref{fig1} indeed shows that the spectra of the SS
galaxies have broad absorption Balmer lines resembling post-starburst
galaxies. This indicates an episode of star formation less than
1 Gyr old \citep[][and refences therein]{P18}.
After the modeling in \S\ref{starlight} and removing the emission of
ionized gas, we determine the equivalent widths (EWs) of
$\mathrm{H}{\alpha}$ to be 
EW($\mathrm{H}{\alpha}$)= $2.2 \pm 0.5$ \AA,
EW($\mathrm{H}{\alpha}$)= $2.3 \pm 0.2$ \AA,
EW($\mathrm{H}{\alpha}$)= $2.9 \pm 0.4$ \AA, and
EW($\mathrm{H}{\alpha}$)= $3.2 \pm 0.9$ \AA in H79a, H79b, H79c, and
H79f, respectively. These values are the same, to within the errors. 
Balmer lines are prominent in B and A stars, whose main-sequence lifetime
\citep[$t_{ms}\sim 10^{10}(M_{\star}/M_{\sun})^{-2.5}$, where
 $M_{\star}$ is the mass for an upper main sequence
 star;][p.28]{HK94} 
runs between 0.1 and 1 Gyr, consistent with the ages enclosed by
region~3 in Fig.~\ref{fig3} where a low level of star formation
continues. The similarity of the EW for the SS galaxies supports
the idea of a recent common history of star formation associated with
the dynamical evolution of the group.
   
There is an important caveat to the qualitative analysis of the sSFH
presented above. The sSFH obtained in Sec.~\ref{sSFH} depend strongly
on the adopted SSP library and the details of the inversion method
\citep[e.g.,][]{Sa19,CF14}. Therefore, our results although consistent
and suggestive, remain quantitively strongly model-dependent.   

\subsection{Kinematics in H79b}\label{rotation}
We analyzed the kinematics of H79b using techniques developed by
\citet{Ba14,GL15}. Kinematic centers for the stellar and ionized gas
components were generated via gradient maps from the velocity fields
shown in Fig.~\ref{fig5}(a). Fig.~\ref{fig5}(b) shows the
pseudo-rotation curves (pRC) \citep{Ba14} for the stellar and 
ionized gas components.  The amplitudes of the pRCs are similar but
counter-rotating.  The maximum stellar rotation speed
in~Fig.~\ref{fig5}(b), $v_{max} \simeq 110\,\mathrm{km\,s^{-1}}$, 
leads to a dynamical mass
$M_{dyn}(r)=1.8\,r \,G^{-1}(0.5v_{max}^2+ \sigma_{gal}^2)$
\citep[][Eq. 5, \S 5.2]{AO18} for H79b within $r=R_{e}$, where
$\sigma_{gal}=145\,{\rm km\,s^{-1}}$ is the velocity dispersion, so
that $M_{dyn}(R_{e}) \eqsim\,2\times 10^{10}M_{\sun}$ (cf., Du08). 

We interpret the counter-rotating disk in H79b as evidence for
multiple encounters with other galaxies and gas removal followed by
re-accretion from the intragroup medium. The numerical simulations of
\citet{St19} support such an origin for counter-rotating disks in
dense environments. \citeauthor{St19} also estimated that
counter-rotating disks could remain stable for 2~Gyr or
more. Therefore, we propose that the counter-rotating disk in H79b was
created during the first crossing (region 2 in Fig~\ref{fig3}).

\subsection{The Distribution of Ionized Gas}\label{spt}
Figure \ref{fig2}(b) maps the distribution of ${\mathrm H}{\alpha}$
in emission above a conservative flux cut of
$1.2\times10^{-16}\; \mathrm{erg\,s^{-1}\,cm^{-2}\,arcsec^{-2}}$.
The ionized gas concentrates towards the galaxy
nuclei. H79c shows only faint emission, while H79f has only fainter
emission. The pattern shown by the IFU data agrees with Fabry-P\'erot
observations by Du08.

\begin{figure}[h]
 \begin{center}
  \plotone{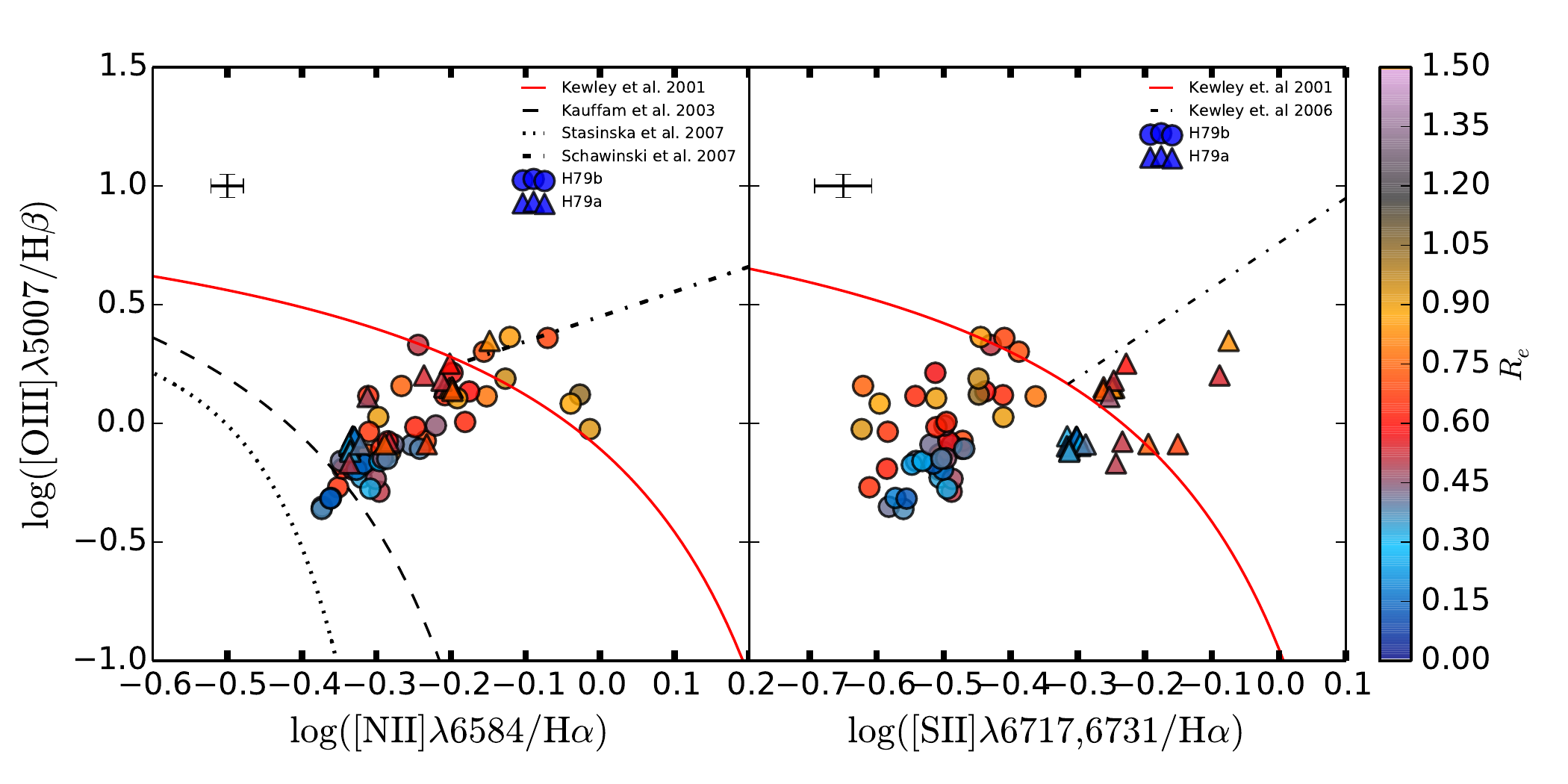}
 \end{center}
 \caption{Line-ratio diagnostic diagrams given by spaxel as a
  function of distance to the center scaled by the effective radius
  ($R_{e}$) for H79a (upright triangles) and H79b (circles). Mean
  errors are indicated on the upper left.  The overplotted black
  lines represent empirically or theoretically derived diagnostics
  for LINER/Seyfert and star formation activity. No spaxel shows
  starburst activity.
 } 
 \label{fig6}
\end{figure}

Line ratios useful for diagnostic diagrams were measured with
uncertainties calculated by propagating the errors in
quadrature. Different classification criteria
\citep{Ka03,Ke01,Kew06,St06,Sc07} 
are overplotted in Figure \ref{fig6}, where the
spaxel radial distribution, scaled by $R_{e}$, is indicated by the
color bar. The distribution for 
H79a (upright colored triangles) shows some AGN/LINER activity, while
H79b is dominated by composite activity (colored circles),
though star formation driven by galaxy-galaxy
interactions \citep[e.g.,][]{Hop09} is seen close to
the nuclei. Nevertheless, no region within the H79a/H79b envelope
has starburst character.

%

The comparison of Fig. \ref{fig6} in this paper with Fig.~6 in
\citet{Sa15} suggests that the underlying stellar populations of H79a
and H79b are old (i.e., with luminosity-weighted age $\gtrsim10^{9.5}$
yrs), but prolonged star-formation activity over the last several Myr
might be present \citep[e.g.,][]{CMM15, Sa15}.  

The results of this subsection are in agreement with \S\ref{sSFH}, as
we expected only a low level of star formation in the SS galaxy
central regions. Moreover, the comparison of our diagnostic diagrams
with earlier results suggest that the  underlying stellar population
is old.

There are indications of biconical outflows in galaxies H79a and H79b,
but we leave the hydrodynamical analysis of the ionized gas for a
further publication.

\subsection{Star Formation derived from IR observations}\label{cmc}
An independent indicator of star formation is provided by IR
properties. We co-added the pure gas spectra (\S\ref{starlight})
across H79b to derive 
${L_{\mathrm{H}{\alpha}}} = (4\pm1)\times10^{40}\; \mathrm{erg\,s^{-1}}$
which leads to the estimate 
$\mathrm{SFR_{\mathrm{H}_\alpha}} = (7.9\times 10^{-42})\times{L_{\mathrm{H}{\alpha}}}
 = 0.3\pm0.1\; \mathrm{M_{\sun}\, yr^{-1}}$ 
\citep{Ke98}. Using the IR luminosity (\S\ref{dust}), we get
$\mathrm{SFR_{IR}}=1.71\times10^{-10}\,L_{\mathrm{IR}}=0.5\pm0.1\,
 \mathrm{M_{\sun}\,yr^{-1}}$ 
\citep{Ke98}. Although the inferred SFR from the dust emissivity is
only accurate to $10\,\%$ \citep{Ca12}, it is encouraging that the two
estimates are in fair agreement. The $\mathrm{H}{\alpha}$ emission  is
not completely sampled across H79a and is low in H79c, hence we have
only calculated the SFR for these galaxies based on the IR scaling
law. We found $\mathrm{SFR_{IR}} = 0.64\pm0.06\,\mathrm{M_{\sun}\,yr^{-1}}$ 
and $0.07 \pm0.02\,\mathrm{M_{\sun}\,yr^{-1}}$, for H79a and H79c,
respectively.  Our SFRs are close to the results of \citet{Bi11}, but
are an order of magnitude higher than \citet{Bi14} --- it is known that
{\tt cmcirsed} underestimates $M_{\mathrm{dust}}$ compared with the
more detailed modeling provided by {\tt MAGPHYS} over a much wider
spectral range \citep{Bi14}. Overall, the discrepancies among the
estimated parameters can be accounted for by the differences in the
definition of SFR and dust attenuation employed in the  
modeling \citep[e.g.,][]{Hu16, Ht19}, and the assumed cosmology.  

In spite of the discrepancies cited above, we conclude that star
formation in H79a, H79b, H79c, and H79f is presently low (cf.,
\S\ref{sSFH}, \S\ref{spt}).  

The  HI distribution  over  SS  also supports the low level of star formation suggested above.  
HI is not scattered over the whole group, it's actually  centered on H79d, showing weak emission extending to the East and NE, which might suggest that H79d is entering SS for the first time (e.g., Du08).  However, HI is deficient  corresponding  to 30-70\% of the expected HI for the optical luminosities and morphological types \citep{VM01, BYV10}. In this work, we  suggest that  SS is going through its second crossing; hence,  the observed HI 
deficiency resulted because HI might have been  exhausted during the first crossing (\S\ref{sSFH}).


\section{Conclusions}\label{con}
\begin{itemize}

\item The analysis presented above allows us to reject the ideas that SS is
 a chance projection or assembling for the first time: we suggest that SS
 is experiencing secondary infall.  The underlying stellar
 population is old and current star formation is low
 (\S\ref{sSFH}, \S\ref{spt}). The distribution of morphologies and the
 compactness of SS galaxies (\S\ref{surface}),  the presence of a
 disrupted galaxy (H79f, \S\ref{sSFH}), displaced HI 
 \citep[e.g., Du08,][]{BYV10}, and luminous intragroup light
 are suggestive of strong interactions.

\item Using a simple dynamical framework based on secondary infall
 in \S\ref{model} we have been able to calculate the first and second
 turn-around radii for SS. We found 
 $r_{1t}\approx 1670 \,h_{75}^{-1}\;\mathrm{kpc}$ and 
 $r_{2t}\approx  530 \,h_{75}^{-1}\;\mathrm{kpc}$.  We also derived a
 virial mass
 $M_{v}\approx8.2\times10^{12}\;h^{-1}_{75}\;M_{\sun}$. Since
 CG are neither expanding nor contracting at the turn-around radius, 
 we can calculate the corresponding free-fall times, and find
 $t_{1t} \sim 11$~Gyr and $t_{2t} \sim 2$~Gyr. Hence, SS is able to
 cross at least once during a Hubble time.  
 
\item We suggest that by using the fossil record of star formation of
 CG galaxies (\S\ref{starlight}, \S\ref{sSFH}), we can identify the
 corresponding crossings with tidally-induced coeval bursts of star
 formation. We generated the sSFHs of the SS galaxies over three
 ranges of radii using P\textsc{ipe}3D, and found
 coordinated episodes of star formation in H79a, H79b, H79c, and H79f
 that are not present in a control sample. Our qualitative analysis
 allow us to identify three episodes of star formation. The first, 
 at stellar age $t_{sa} \gtrsim 10$ Gyrs ($z \gtrsim 2.3$),
 corresponds to the formation of each galaxy. The second episode
 times the first crossing to stellar age  
 $t_{sa} \thicksim 3$~Gyrs ($z_{fc} \thicksim 0.3$) and lasted for
 about 3.5 Gyrs. We found an ongoing episode of mild star formation at
 $t_{sa}=0.02$ Gyr, from which we suggest that the SS galaxies began
 their second crossing around $z_{sc}\thicksim 0.016$. This may be the
 origin of the extreme compactness of SS.
 
\item The presence and stability (for more than $\gtrsim 2$ Gyr) of a
 counter-rotating disk in H79b provides an independent timing for the
 first crossing (\S\ref{rotation}).  

\item The integrated spectra of H79a, H79b, H79c and H79f
 resemble those of post-starburst galaxies (Fig.~\ref{fig1}). This is
 indicative of a recent episode of star formation not more than
 1~Gyr ago. A recent coordinated episode of star formation would
 result in strong Balmer lines having the same EW. Indeed, we found
 that the four galaxies have an EW 
 $\sim 2.6\,\mathrm{\AA}$ in $\mathrm{H} \alpha$ (\S\ref{sSFH}). 
 This episode of star formation could be $0.1 - 1$~Gyr old, giving
 another bound on the time of the second crossing, independent of the
 adopted stellar population model.
 
\item The analysis  of  diagnostic diagrams (\S\ref{spt}) and the IR
 emission (\S\ref{cmc}) of SS galaxies suggest that star
 formation is currently at a low level throughout the group.
 
\item The complex history of galaxy interactions suggested by our
 dynamical model (\S\ref{model}) and the fossil record (\S\ref{sSFH})
 explains why mergers in CG are dry or mixed and the origin and
 enrichment of the intragroup medium.  

\item The techniques and dynamical scenario used in this paper may
 help to explore the assembly of groups and clusters of galaxies.  

\item The scenario that we have advanced in this paper  provides a
  plausible solution to the short crossing-time paradox in CG, and may
  serve to constrain other models \citep[e.g.,][]{Ath97,TA17}.   
\end{itemize}

\acknowledgments
We acknowledge the careful revision of our paper and most relevant
suggestions of an anonymous referee, which helped us to greatly improve
the results and discussions presented above. We are very grateful to
Prof.~Frederic A. Rasio, the Letters Editor, for allowing us to
exceed the nominal length limits to answer to the referee's remarks
and suggestions. OLC acknowledges  enlightening discussions with
Marina Rodr{\'\i}guez-Baras, Luis Aguilar, Josh Barnes,  
Theo Bitsakis, H\'ector Aceves  and Gustavo Bruzual.  
SFS thanks PAPIIT-DGAPA-IN100519, Conacyt CB-285080, and
FC-2016-01-1916 projects. This study uses data from the CALIFA
survey\footnote{\url{http://califa.caha.es/}}. Based on observations
collected at CAHA, operated jointly by MPIA and  IAA (CSIC).



\facility{CAO:3.5 (PMAS,PPak), HST (WFPC2), WISE)}, 




\begin{thebibliography}{}

\bibitem[Aquino-Ort{\'{\i}}z et al.(2018)]{AO18} Aquino-Ort{\'{\i}}z, E., Valenzuela, O., S{\'a}nchez, S.~F., et al.\ 2018, \mnras, 479, 2133 

\bibitem[Athanassoula et al.(1997)]{Ath97} Athanassoula, E., Makino, J., \& Bosma, A.\ 1997, \mnras, 286, 825

\bibitem[Barrera-Ballesteros et 
al.(2014)]{Ba14} Barrera-Ballesteros, J.~K., Falc{\'o}n-Barroso, J., Garc{\'{\i}}a-Lorenzo, B., et al.\ 2014, \aap, 568, A70  

\bibitem[Barnes(1985)]{Ba85} Barnes, J.\ 1985, \mnras, 215, 517.

\bibitem[Barnes(1997)]{Ba97} Barnes, J.~E.\ 1997, The Nature of
  Elliptical Galaxies; 2nd Stromlo Symposium, 116, 469  

\bibitem[Bertschinger(1985)]{Be85} Bertschinger, E.\ 1985, 
\apjs, 58, 39 

\bibitem[Bitsakis et al.(2011)]{Bi11} Bitsakis, T., Charmandaris, V., da Cunha, E., et al.\ 2011, \aap, 533, A142.

\bibitem[Bitsakis et al.(2014)]{Bi14} Bitsakis, T., Charmandaris, V., Appleton, P.~N., et al.\ 2014, \aap, 565, A25 

\bibitem[Borthakur et al.(2010)]{BYV10} Borthakur, S., Yun,  M.~S., \& Verdes-Montenegro, L.\ 2010, \apj, 710, 385 


\bibitem[Cappellari \& Emsellem(2004)]{CE04} Cappellari, M., \& Emsellem, E.\ 2004, \pasp, 116, 138

\bibitem[Cardelli et al.(1989)]{Ca89} Cardelli, J.~A., Clayton, G.~C., \& Mathis, J.~S.\ 1989, \apj, 345, 245 

\bibitem[Casey(2012)]{Ca12} Casey, C.~M.\ 2012, \mnras, 425, 
3094


\bibitem[Cid Fernandes et al.(2005)]{CF05} Cid Fernandes, R., Mateus, A., Sodr{\'e}, L., Stasi{\'n}ska, G., \& Gomes, J.~M.\ 2005, \mnras, 358, 363


\bibitem[Cid Fernandes et al.(2014)]{CF14} Cid Fernandes, R., Gonz{\'a}lez Delgado, R.~M., Garc{\'{\i}}a Benito, R., et al.\ 2014, \aap, 561, A130 

\bibitem[Cid Fernandes et al.(2015)]{CF15} Cid Fernandes, R., Lacerda, E.~A.~D., Gonz{\'a}lez Delgado, R.~M., et al.\ 2015, Galaxies in 3D Across the Universe, 93


\bibitem[Coenda et al.(2015)]{CMM15} Coenda, V., Muriel, H., \& Mart{\'{\i}}nez, H.~J.\ 2015, \aap, 573, A96 

\bibitem[Conroy(2013)]{Co13} Conroy, C.\ 2013, \araa, 51, 393


\bibitem[Courteau et al.(2014)]{Co14} Courteau, S., Cappellari, M., de Jong, R.~S., et al.\ 2014, Reviews of Modern Physics, 86, 47


\bibitem[Da Rocha \& Mendes de Oliveira(2005)]{DR05} Da Rocha, C., \& Mendes de Oliveira, C.\ 2005, \mnras, 364, 1069

\bibitem[Diaferio et al.(1993)]{Di93} Diaferio, A., Ramella, M., Geller, M.~J., \& Ferrari, A.\ 1993, \aj, 105, 2035 

\bibitem[D{\'{\i}}az-Gim{\'e}nez \& Zandivarez(2015)]{DZ15} D{\'{\i}}az-Gim{\'e}nez, E., \& Zandivarez, A.\ 2015, \aap, 578, A61 

\bibitem[D'Souza, \& Bell(2018)]{DS18} D'Souza, R., \& Bell, E.~F.\ 2018, Nature Astronomy, 2, 737

\bibitem[Durbala et al.(2008)]{Dur08} Durbala, A., del Olmo, 
A., Yun, M.~S., et al.\ 2008, \aj, 135, 130 [Du08]

\bibitem[Garc{\'\i}a-Benito et al.(2015)]{GB15} Garc{\'\i}a-Benito, R., Zibetti, S., S{\'a}nchez, S.~F., et al.\ 2015, \aap, 576, A135


\bibitem[Garc{\'{\i}}a-Lorenzo et al.(2015)]{GL15} Garc{\'{\i}}a-Lorenzo, B., M{\'a}rquez, I., Barrera-Ballesteros, J.~K., et al.\ 2015, \aap, 573, A59 

\bibitem[Gonz{\'a}lez Delgado et al.(2014)]{GD14} Gonz{\'a}lez Delgado, R.~M., Cid Fernandes, R., Garc{\'\i}a-Benito, R., et al.\ 2014, \apjl, 791, L16

\bibitem[Hansen \& Kawaler(1994)]{HK94} Hansen, C.~J., \& Kawaler, S.~D.\ 1994, Stellar Interiors. Physical Principles


\bibitem[Hickson(1993)]{H93} Hickson, P.\ 1993, Astrophysical Letters and Communications, 29, 1

\bibitem[Hopkins et al.(2009)]{Hop09} Hopkins, P.~F., Cox, T.~J., Younger, J.~D., \& Hernquist, L.\ 2009, \apj, 691, 1168 

\bibitem[Hung et al.(2016)]{Hu16} Hung, C.-L., Casey, C.~M., Chiang, Y.-K., et al.\ 2016, \apj, 826, 130

\bibitem[Hunt et al.(2019)]{Ht19} Hunt, L.~K., De Looze, I., Boquien, M., et al.\ 2019, \aap, 621, A51

\bibitem[Ibarra-Medel et al.(2016)]{Ib16} Ibarra-Medel, H.~J., S{\'a}nchez, S.~F., Avila-Reese, V., et al.\ 2016, \mnras, 463, 2799 

\bibitem[Ibarra-Medel et al.(2019)]{Ib19} Ibarra-Medel, H.~J., Avila-Reese, V., S{\'a}nchez, S.~F., et al.\ 2019, \mnras, 483, 4525


\bibitem[Kauffmann et al.(2003)]{Ka03} Kauffmann, G., Heckman, T.~M., Tremonti, C., et al.\ 2003, \mnras, 346, 1055 

\bibitem[Kelz et al.(2006)]{Ke06} Kelz, A., Verheijen, M.~A.~W., Roth, M.~M., et al.\ 2006, \pasp, 118, 129

\bibitem[Kennicutt(1998)]{Ke98} Kennicutt, R.~C., Jr.\ 1998, \araa, 36, 189 


\bibitem[Kewley et al.(2001)]{Ke01} Kewley, L.~J., Dopita, 
M.~A., Sutherland, R.~S., Heisler, C.~A., \& Trevena, J.\ 2001, \apj, 556, 121 

\bibitem[Kewley et al.(2006)]{Kew06} Kewley, L.~J., Groves, 
B., Kauffmann, G., \& Heckman, T.\ 2006, \mnras, 372, 961

\bibitem[Lynden-Bell(1981)]{LB81} Lynden-Bell, D.\ 1981, The Observatory, 101, 111

\bibitem[Lynden-Bell(1994)]{LB94} Lynden-Bell, D.\ 1994, The Formation and Evolution of Galaxies, 85

\bibitem[Lynden-Bell(1999)]{LB99} Lynden-Bell, D.\ 1999, The Stellar Content of Local Group Galaxies, 39

\bibitem[Nishiura et al.(2002)]{Ni02} Nishiura, S., Shioya, Y., Murayama, T., et al.\ 2002, \pasj, 54, 21 


\bibitem[Pawlik et al.(2018)]{P18} Pawlik, M.~M., Taj Aldeen, L., Wild, V., et al.\ 2018, \mnras, 477, 1708


\bibitem[Peebles(1971)]{Pe71} Peebles, P.~J.~E.\ 1971, Physical Cosmology, Princeton Series in Physics, Princeton, N.J.: Princeton University Press, 1971, pp. 83-86.

\bibitem[Peng et al.(2010)]{Pe10} Peng, C.~Y., Ho, L.~C., Impey, C.~D., \& Rix, H.-W.\ 2010, \aj, 139, 2097

\bibitem[P{\'e}rez et al.(2013)]{Pe13} P{\'e}rez, E., Cid Fernandes, R., Gonz{\'a}lez Delgado, R.~M., et al.\ 2013, \apjl, 764, L1 


\bibitem[Plauchu-Frayn et al.(2012)]{PF12} Plauchu-Frayn, I., Del Olmo, A., Coziol, R., et al.\ 2012, \aap, 546, A48.


\bibitem[Roth et al.(2005)]{Ro05} Roth, M.~M., Kelz, A., Fechner, T., et al.\ 2005, \pasp, 117, 620

\bibitem[S{\'a}nchez et al.(2015)]{Sa15} S{\'a}nchez, S.~F., P{\'e}rez, E., Rosales-Ortega, F.~F., et al.\ 2015, \aap, 574, A47 

\bibitem[S{\'a}nchez et al.(2016)]{Sa16a} S{\'a}nchez, S.~F., Garc{\'{\i}}a-Benito, R., Zibetti, S., et al.\ 2016, \aap, 594, A36

\bibitem[S{\'a}nchez et al.(2016a)]{Sa16b} S{\'a}nchez, S.~F., P{\'e}rez, E., S{\'a}nchez-Bl{\'a}zquez, P., et al.\ 2016a, \rmxaa, 52, 21 

\bibitem[S{\'a}nchez et al.(2016b)]{Sa16c} S{\'a}nchez, S.~F., P{\'e}rez, E., S{\'a}nchez-Bl{\'a}zquez, P., et al.\ 2016b, \rmxaa, 52, 171 
 
 \bibitem[S{\'a}nchez et al.(2019)]{Sa19} S{\'a}nchez, S.~F., Avila-Reese, V., Rodr{\'{\i}}guez-Puebla, A., et al.\ 2019, \mnras, 482, 1557 


\bibitem[Schawinski et al.(2007)]{Sc07} Schawinski, K., Thomas, D., Sarzi, M., et al.\ 2007, \mnras, 382, 1415 

\bibitem[Schlegel et al.(1998)]{Sch98} Schlegel, D.~J., Finkbeiner, D.~P., \& Davis, M.\ 1998, \apj, 500, 525

\bibitem[Seyfert(1951)]{Se51} Seyfert, C.~K.\ 1951, \pasp, 63, 72

\bibitem[Silk, et al.(2014)]{SCD14} Silk, J., Di Cintio, A., \& Dvorkin, I.\ 2014, Proceedings of the International School of Physics 'Enrico Fermi' Course 186 'New Horizons for Observational Cosmology' Vol. 186, 137

\bibitem[Starkenburg et al.(2019)]{St19} Starkenburg, T.~K., Sales, L.~V., Genel, S., et al.\ 2019, arXiv e-prints , arXiv:1903.03627.


\bibitem[Stasi{\'n}ska et al.(2006)]{St06} Stasi{\'n}ska, G., Cid Fernandes, R., Mateus, A., Sodr{\'e}, L., \& Asari, N.~V.\ 2006, \mnras, 371, 972


\bibitem[Tamayo \& Aceves(2017)]{TA17} Tamayo, F.~J., \& Aceves, H.\ 2017, \rmxaa, 53, 515 

\bibitem[Thomas et al.(2010)]{TM10} Thomas, D., Maraston, C., Schawinski, K., et al.\ 2010, \mnras, 404, 1775

\bibitem[Tully(2015)]{Tu14} Tully, R.~B.\ 2015, \aj, 149, 54, [Tu15] 

\bibitem[van de Weygaert \& Bond(2008)]{vdWB08} van de Weygaert, R., \& Bond, J.~R.\ 2008, A Pan-Chromatic View of Clusters of Galaxies and the Large-Scale Structure, 740, 335 

\bibitem[Verdes-Montenegro et al.(2001)]{VM01} Verdes-Montenegro, L., Yun, M.~S., Williams, B.~A., et al.\ 2001, \aap, 377, 812

\bibitem[Walcher et al.(2011)]{Wa11} Walcher, J., Groves, B., Budav{\'a}ri, T., et al.\ 2011, \apss, 331, 1

\bibitem[White(1990)]{W89} White, S.~D.~M.\ 1990, Dynamics and Interactions of Galaxies, Rolan Wielen editor, Springer-Verlag,  Berlin Heidelberg, 380 

\bibitem[Zemcov et al.(2014)]{Zem14} Zemcov, M., Smidt, J., 
Arai, T., et al.\ 2014, Science, 346, 732 


\end{thebibliography}
\end{document}